# Oxygen-annealing effects on superconductivity in polycrystalline $Fe_{1-x}Te_{1-y}Se_y$

Gina M. Friederichs, Matthias P. B. Wörsching, Dirk Johrendt*

Department of Chemistry, Ludwig-Maximilians-University Munich, Butenandtstrasse 5-13, 81377 Munich, Germany

E-mail: johrendt@lmu.de

Abstract:

Superconductivity in *anti*-PbO-type iron chalcogenides $Fe_{1+x}Te_{1-y}Se_y$ ($x$ = 0, 0.1, $y$ = 0.1-0.4) depends on the amount ($x$) of interstitial iron atoms located between the $FeTe_{1-y}Se_y$ layers. Non-superconducting samples of nominal $Fe_{1.1}Te_{1-y}Se_y$ convert to superconductors with critical temperatures up to 14 K after annealing at 300°C in an oxygen atmosphere. The process is irreversible upon subsequent hydrogen annealing. Magnetic measurements are consistent with the formation of iron oxides suggesting that oxygen annealing preferably extracts interstitial iron from $Fe_{1+x}Te_{1-y}Se_y$ which interfere with superconductivity.

Introduction:

Superconducting iron chalcogenides Fe*Ch* (*Ch* = Se, Te) attract considerable interest because their simple *anti*-PbO-type crystal structures make them perfect candidates for studying unconventional superconductivity,[1] and because recent reports indicate critical temperatures ($T_c$) up to 100 K in single-layer FeSe films.[2]

Fe*Ch* compounds consist of stacked layers of edge sharing $FeCh_{4/4}$ tetrahedra with up to 25 % iron located in the van der Waals gap.[3] These interstitial iron atoms in $Fe_{1+x}Ch$ interfere with superconductivity by their magnetic moments and/or their unfavourable contributions to the Fermi surface. Virtually stoichiometric FeSe[4] has critical temperatures of 8-10 K[5] which increases to 36 K under pressure[6] and up to 45 K by intercalation of innocent electron transferring species.[7] Stoichiometric FeTe is unknown while $Fe_{1+x}Te$ ($x \approx 0.14$) is magnetic and non-superconducting. The critical temperatures of solid solution $FeTe_{1-y}Se_y$ raises to 14 K at $y \approx 0.5$ depending on the amount of interstitial iron.[8] Thus understanding and controlling the excess iron in Fe*Ch* superconductors is fundamentally important also with respect to possible applications in superconducting wires.

Different post-preparation manipulations revealed significant influences on the superconducting properties of $FeTe_{1-y}Se_y$. These treatments include exposure to $HNO_3$[9], $I_2$[10], $O_2$[11], Te[12] and S[13] at ambient conditions or annealing at higher temperatures as well as under $N_2$[10] or vacuum.[14] Especially annealing $Fe_{1+x}Te_{1-y}Se_y$ single crystals in an oxygen atmosphere improved the superconducting properties significantly. As-prepared samples are superconducting only for $y = 0.5$, whereas after $O_2$-annealing already compounds with $y = 0.1$ are superconducting.[11] Since then, the mechanism of $O_2$-annealing has been debated. Several assumptions including a homogenisation effect, the substitution of O for *Ch*, the intercalation of oxygen or the removal of interstitial Fe, were considered reasonable, whereby the latter is discussed preferentially.[11, 15] Recently the removal of excess iron from a $Fe_{1+x}Te_{0.6}Se_{0.4}$ single crystal was monitored by STM measurements.[16] However, it remains unclear what happens with that iron, and if the process is solely the removal of interstitial iron while the layer iron is unaffected. Maybe it is more complex and possibly reversible under reductive conditions in hydrogen atmosphere.

Here we study polycrystalline samples of $Fe_{1+x}Te_{1-y}Se_y$ ($x = 0, 0.1, y = 0.1$-$0.4$) with different amounts of nominal interstitial iron concerning the influences of oxygen- and hydrogen-annealing on the superconducting properties. In order to exclude effects due to heating we also performed control experiments under Ar-atmosphere. If the extraction of interstitial Fe atoms is essential, the emergence of iron oxide as impurity phase can be expected in polycrystalline samples and the process should be irreversible under reductive conditions.

Methods:

**Synthesis.** $Fe_{1+x}Te_{1-y}Se_y$ ($x = 0$-$0.1$, $y = 0.1$-$0.4$) was synthesized by using stoichiometric amounts of the elements. These samples will be referred to as "as-prepared" in the following. $Fe_{1.0}Te_{1-y}Se_y$ ($y = 0.1$-$0.4$) samples (0.7 g) were synthesized in alumina crucibles inside sealed silica ampoules by heating to 1050 °C for 24 h, cooled to 350 °C for 10 h and then cooled to room temperature (step 1). Four samples were combined and annealed at 800 °C for 10 h followed by 10 h at 350 °C before cooling to room temperature (step 2). $Fe_{1.1}Te_{1-y}Se_y$ ($y = 0.1$-$0.4$) was synthesized in one step according to step 1 but on a larger scale (2.0 g). Oxygen annealing was performed by heating samples to 300 °C for 2 h in alumina crucibles inside sealed Duran© glass ampoules under oxygen atmosphere ("$O_2$-annealed" samples). For hydrogen annealing, $O_2$ treated samples in alumina crucibles inside a Duran tube connected to a bubble counter were heated to 200 °C for 2 h under a continuous flow of hydrogen ("H$_2$-annealed" samples). Control experiments were performed with as-prepared samples under Ar atmosphere at 300 °C for 2 h.

**X-ray Powder Diffraction.** X-ray powder diffraction patterns were recorded on a STOE Stadi P (MoKα1 radiation, Ge (111) primary monochromator, λ = 70.93 pm, silicon as external standard, rotating capillary 0.3 mm outer diameter). Rietveld refinements were performed with the TOPAS 4.1 program package.[17] To generate the reflection profiles the fundamental parameters approach was used. The preferred orientation of the crystallites was described with fourth-order spherical harmonics.

**EDX.** Scanning electron microscopy of polycrystalline samples was performed on a Carl Zeiss EVO-MA 10 microscope with SE and BSE detectors, which was controlled by the SmartSEM[18] software. The microscope was equipped with a Bruker Nano EDS detector (X-Flash detector 410-M) for EDS investigations using the QUANTAX 200[19] software to collect and evaluate the spectra. Elements contained in the sample holder and adhesive carbon tabs were disregarded.

**Magnetic Measurements.** A commercial Quantum Design MPMS XL5 SQUID magnetometer was used at temperatures between of 1.8 and 300 K. The polycrystalline sample was ground and filled into a gelatin capsule, which was fixed in a plastic straw. The magnetic measurements were performed with the MPMS MultiVu software. AC-susceptibility measurements were performed on a fully automatic self-built AC-susceptometer with a Janis SHI-950 two-stage closed-cycle Cryostate with $^4$He exchange gas (Janis Research Company, Wilmington, U.S.A.) and a dual-channel temperature controller (model 332 by LakeShore, Westerville, U.S.A.) at temperatures between 3.5 and 300 K.

**Conductivity Measurements.** The electrical measurements were also performed with the self-built susceptometer. A Keithley Source- Meter 2400 (Cleveland, U.S.A.) was available as current source. The differential voltage drop between signal-high and signal-low was recorded with a Keithley 2182 Nano-Voltmeter and used to calculate the sample resistance in one direction according to Ohm's law and the specific resistance according to the Van-der-Pauw approximation. For the measurements cold pressed (5 kN) pellets of respective samples (diameter, 4.0 mm; thickness, 0.4−0.9 mm) were produced. Applying the four-probe method, the pellet was contacted with four equidistant probes using silver conducting paint. All preparations were performed under inert atmosphere in a glovebox.

Results and Discussion:

As-prepared, $O_2$-annealed and $H_2$-annealed polycrystalline $FeTe_{1-y}Se_y$ and $Fe_{1.1}Te_{1-y}Se_y$ samples were characterized by x-ray powder diffraction (XRPD) and Rietveld refinements (**Figure 1**). An impurity of $FeTe_2$ ($\leq$ 12 %) occurs in all $FeTe_{1-y}Se_y$ samples except in as-prepared $Fe_{1.1}Te_{1-y}Se_y$. Comparing as-prepared with $O_2$- and $H_2$-annealed samples reveals minor changing XRPD intensity ratios along with an increase ($FeTe_{1-y}Se_y$) or evolution ($Fe_{1.1}Te_{1-y}Se_y$, $y$ = 0.1, 0.2) of $FeTe_2$ (asterisks, **Figure 1**). Crystallographic details are given in **Table SI1**.

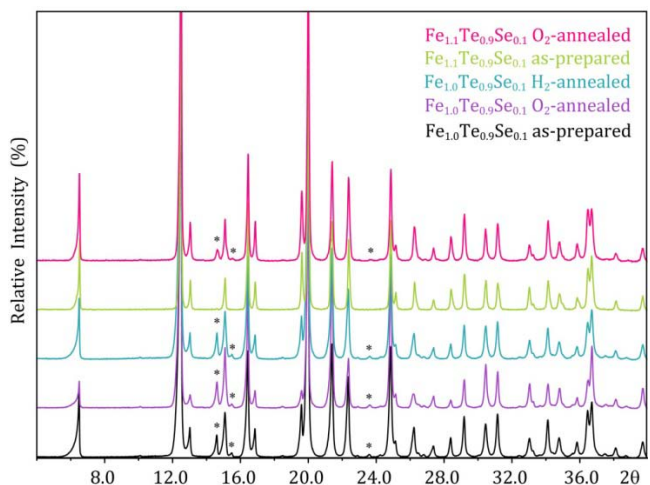

**Figure 1.** XRPD patterns of $Fe_{1+x}Te_{0.9}Se_{0.1}$ as prepared, $O_2$-annealed and $H_2$-annealed.

The lattice parameters $a$ decrease slightly by 0.4 % with increasing Se content $y$, while $c$ decrease stronger by 2.7 %, as expected. After annealing only small changes not exceeding 0.1 % were found.

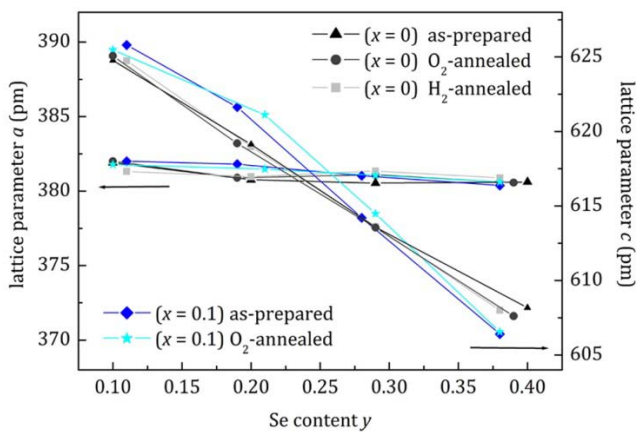

**Figure 2.** Lattice parameters of $Fe_{1+x}Te_{1-y}Se_y$ ($x$ = 0, 0.1, $y$ = 0.1-0.4).

Small changes of interstitial iron amounts are unreliably detectable by XRPD because of very weak scattering of the only ≈ 10 % Fe2 occupied $2c$ site (**Figure SI1**). It is furthermore likely that the oxygen treatment proceeds from the surface leading to inhomogeneous particles.[16] However the structural data from the Rietveld refinements agree with those obtained by single crystal methods ($d$(Fe1-Se) ≈ 240 pm; $d$(Fe1-Te) ≈ 260 pm; chalcogen heights: $h_{Se}$ ≈ 150 pm; $h_{Te}$ ≈ 170 pm).[20]

The compositions were determined by careful EDX analysis. Iron contents in **Table 1** are relative to the sums of Te and Se. All ratios Te : Se agree well with the nominal composition and are constrained to 1. The iron contents of O$_2$-annealed samples are slightly reduced while selenium and tellurium contents are unaffected. Most significant is the large amount of 20-30% oxygen in all phases after O$_2$-annealing. This indicates oxide species at the surfaces of the particles. If these are at least partially iron oxides, the iron content inside the O$_2$-annealed particles is lower than given in Table 1 because EDX cannot discriminate iron in the surface oxide and in $Fe_{1+x}Te_{1-y}Se_y$.

Forcing the reaction in a continuous oxygen flow for 2 h at 300 °C instead of static O$_2$ pressure in sealed ampoules partly decomposes $FeTe_{1-y}Se_y$ to impurity phases, among them FeTe$_2$ and iron oxides discernible in powder XRPD. This agrees with the recent results by *Sun et al.* who over-annealed single crystals of $Fe_{1+x}Te_{0.6}Se_{0.4}$ at 400 °C.[16]

**Table 1.** Sample compositions from EDX analysis

| Nominal | as prepared | O$_2$-annealed |
|---|---|---|
| FeTe$_{0.9}$Se$_{0.1}$ | Fe$_{1.03}$Te$_{0.90}$Se$_{0.10}$ (O$_{0.01}$) | Fe$_{1.01}$Te$_{0.90}$Se$_{0.10}$(O$_{0.24}$) |
| FeTe$_{0.8}$Se$_{0.2}$ | Fe$_{1.07}$Te$_{0.80}$Se$_{0.20}$ (O$_{0.02}$) | Fe$_{1.02}$Te$_{0.81}$Se$_{0.19}$(O$_{0.20}$) |
| FeTe$_{0.7}$Se$_{0.3}$ | Fe$_{1.03}$Te$_{0.71}$Se$_{0.29}$ (O$_{0.04}$) | Fe$_{0.99}$Te$_{0.71}$Se$_{0.29}$ (O$_{0.35}$) |
| FeTe$_{0.6}$Se$_{0.4}$ | Fe$_{1.06}$Te$_{0.60}$Se$_{0.40}$ (O$_{0.02}$) | Fe$_{1.02}$Te$_{0.61}$Se$_{0.39}$ (O$_{0.19}$) |
| Fe$_{1.1}$Te$_{0.9}$Se$_{0.1}$ | Fe$_{1.09}$Te$_{0.89}$Se$_{0.11}$ (O$_{0.02}$) | Fe$_{1.05}$Te$_{0.90}$Se$_{0.10}$(O$_{0.32}$) |
| Fe$_{1.1}$Te$_{0.8}$Se$_{0.2}$ | Fe$_{1.11}$Te$_{0.81}$Se$_{0.19}$ (O$_{0.02}$) | Fe$_{1.07}$Te$_{0.79}$Se$_{0.21}$(O$_{0.25}$) |
| Fe$_{1.1}$Te$_{0.7}$Se$_{0.3}$ | Fe$_{1.10}$Te$_{0.72}$Se$_{0.28}$ (O$_{0.01}$) | Fe$_{1.08}$Te$_{0.71}$Se$_{0.29}$(O$_{0.20}$) |
| Fe$_{1.1}$Te$_{0.6}$Se$_{0.4}$ | Fe$_{1.11}$Te$_{0.62}$Se$_{0.38}$ (O$_{0.01}$) | Fe$_{1.08}$Te$_{0.62}$Se$_{0.38}$(O$_{0.21}$) |

As-prepared, O$_2$-, H$_2$- and Ar-annealed samples of FeTe$_{1-y}$Se$_y$ are superconducting with $T_c$ up to 14 K according to AC-susceptibility data shown in **Figure 3a**. In contrast, as-prepared Fe$_{1.1}$Te$_{1-y}$Se$_y$ samples are not superconducting and convert to superconductors only after O$_2$-annealing (**Figure 3b**). Oxygen treatments significantly shift the superconducting transitions to lower selenium con-centrations. As-prepared $FeTe_{0.9}Se_{0.1}$ is not superconducting until $O_2$-annealing induces a $T_c$ of 12 K.

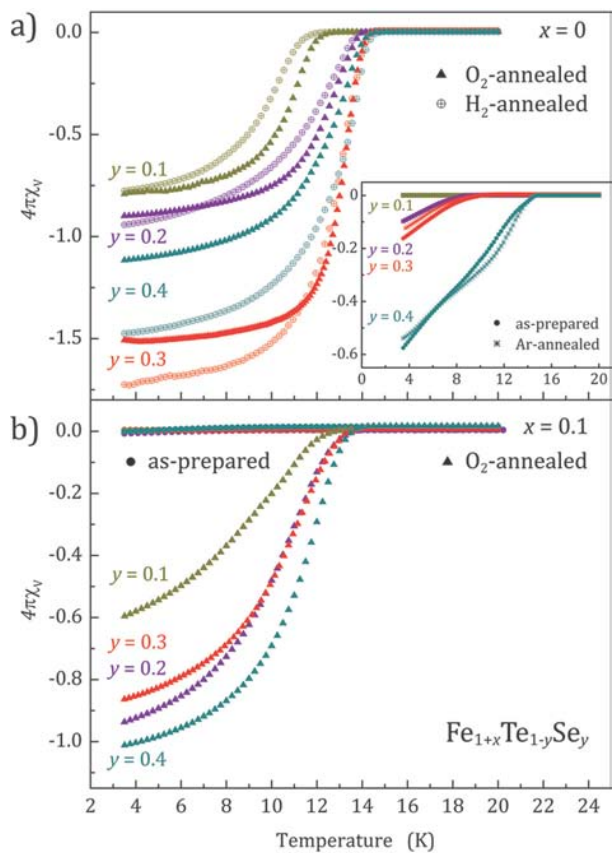

**Figure 3.** AC magnetic susceptibilities of (a) $FeTe_{1-y}Se_y$ after $O_2$- and $H_2$-annealing (Insert: as-prepared and Ar-annealed); (b) $Fe_{1.1}Te_{1-y}Se_y$ as-prepared and after $O_2$-annealing.

Electrical transport data are in line with these findings. The resistivity of as-prepared, $O_2$-annealed and $H_2$-post-annealed $FeTe_{0.8}Se_{0.2}$ drop to zero at 14 K (**Figure 4**). On the contrary, $Fe_{1.1}Te_{0.8}Se_{0.2}$ is only superconducting after $O_2$-treatment (Insert in **Figure 4**).

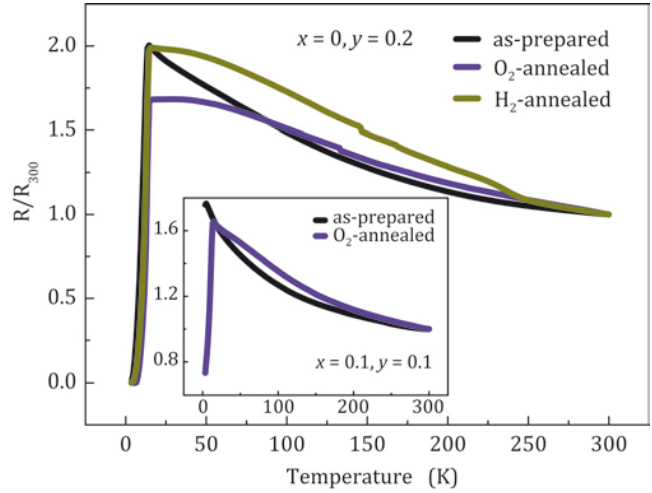

**Figure 4.** Relative electrical resistivities of as-prepared, $O_2$-annealed and $H_2$-annealed FeTe$_{0.8}$Se$_{0.2}$.

The oxygen treatment strongly affects the high field DC magnetization. The magnetic moment ($\mu/\mu_B$) of as-prepared superconducting FeTe$_{0.8}$Se$_{0.2}$ is weak and increases linearly, while a clear S-shaped ferromagnetic background occurs after $O_2$-annealing (Insert in **Figure 5**). This effect is even stronger at 1.8 K, where we find the superposition of a ferromagnetic hysteresis with the magnetization of a type-II superconductor (**Figure 5**). The magnetization at 1.8 K at highest external field remains very weak ($\leq 0.05\ \mu_B$) which means that the ferromagnetism is not a bulk property but caused by a magnetic impurity phase.

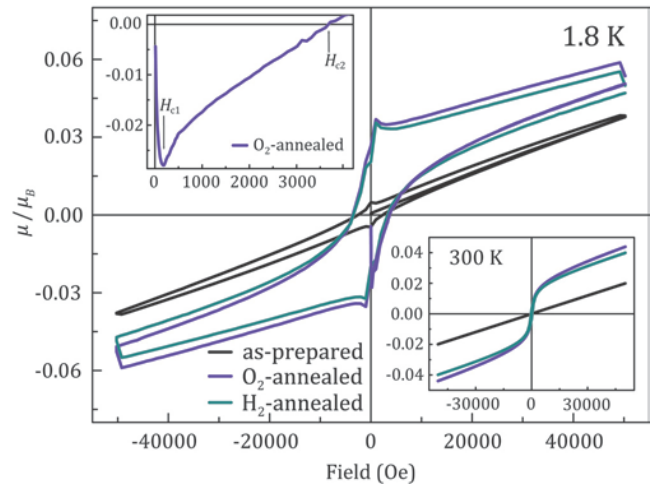

**Figure 5.** Isothermal magnetization of as prepared, $O_2$- and $H_2$-annealed FeTe$_{0.8}$Se$_{0.2}$ at 1.8 K, (Insert: at 300 K, left insert: initial curve of $O_2$-annealed FeTe$_{0.8}$Se$_{0.2}$.)

From our data we estimate that already 0.5 % $Fe_3O_4$ ($T_C$ = 858 K, $\mu$ = 4.1 $\mu_B$) or 0.8 % $\gamma$-$Fe_2O_3$ ($T_C$ = 948 K, $\mu$ = 2.5 $\mu_B$) would produce these magnetization.[21] Such tiny amounts are certainly undetectable by XRPD, which strongly supports the idea that $O_2$-annealing extracts iron from the $Fe_{1+x}Te_{1-y}Se_y$ compounds and forms iron oxides that probably reside at the surface of the particles. The emerging ferromagnetic contribution is furthermore clearly visible in magnetic susceptibility measurements measured under zero field cooled (zfc) and (fc) conditions depicted in **Figure 6**. After annealing the curves are significantly shifted to positive susceptibilities.

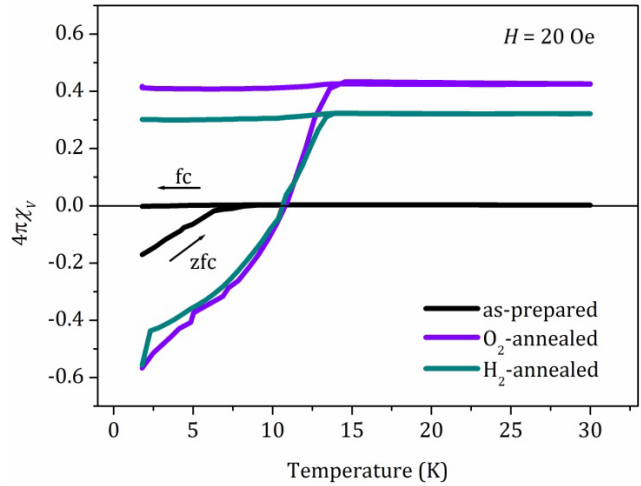

**Figure 6.** Magnetic susceptibility under zero field cooled (zfc) and field cooled (fc) conditions of as-prepared (black), $O_2$-annealed (purple) and $H_2$-annealed (green) $FeTe_{0.8}Se_{0.2}$.

Conclusion:

By combining all findings we conclude that oxygen treatment at 300 °C improves the superconducting properties of polycrystalline $Fe_{1+x}Te_{1-y}Se_y$ through irreversible oxidative de-intercalation of interstitial iron atoms. Traces of magnetic iron oxides are formed. The heterogeneous reaction begins at the surface and probably causes inhomogeneous particles accompanied by $FeTe_2$ impurity formation. Thus the *anti*-PbO-type phase obviously rather degrades if iron is extracted from the layers of $Fe(Te_{1-y}Se_y)_{4/4}$ tetrahedra, in contrast to the formation of iron-deficient layers in $K_{1-x}Fe_{2-y}Se_2$[22] or in $Na_{1-x}Fe_{2-y}As_2$ by sodium de-intercalation of NaFeAs at comparable mild reaction conditions.[23]

Supporting Information:

Interstitial iron content of $Fe_{1+x}Te_{1-y}Se_y$ from Rietveld refinements; Refined structural parameters, exemplarily for $FeTe_{0.9}Se_{0.1}$.

Acknowledgment:

This work was financially supported by the EU collaborative project SUPER-IRON (grant No. 283204)

# Oxygen-annealing effects on superconductivity in polycrystalline $Fe_{1-x}Te_{1-y}Se_y$

Gina M. Friederichs, Matthias P. B. Wörsching, Dirk Johrendt*

Department of Chemistry, Ludwig-Maximilians-University Munich, Butenandtstrasse 5-13 (D), 81377 Munich, Germany

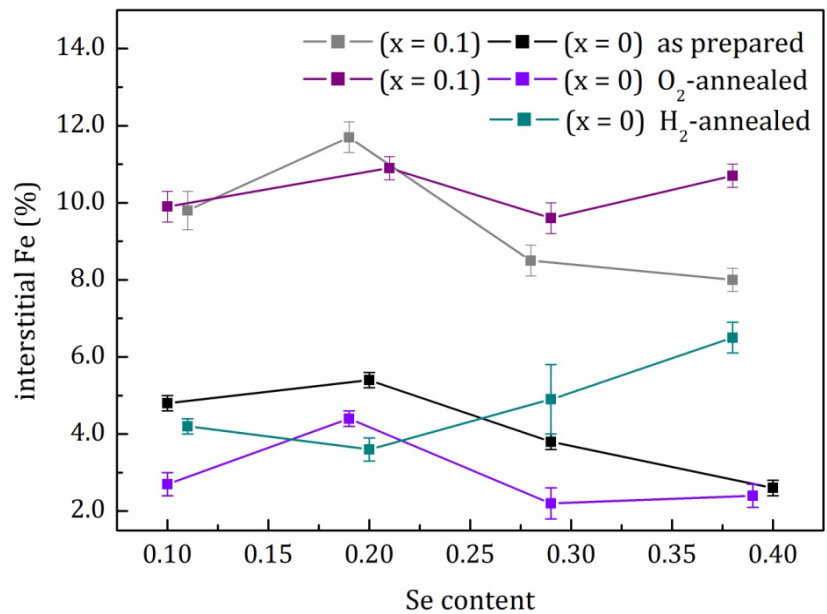

Figure S1. Interstitial Fe content in $Fe_{1+x}Te_{1-y}Se_y$ ($x$ = 0-0.1, $y$ = 0.1-0.4) refined from powder XRD data.

Table SI1. Refined structural parameters for $FeTe_{0.9}Se_{0.1}$.

| Empirical formula | $FeTe_{0.9}Se_{0.1}$ | | |
|---|---|---|---|
| | as prepared | $O_2$- annealed | $H_2$- annealed |
| Refined comp. | $Fe_{1.05(1)}Te_{0.91(1)}Se_{0.09(1)}$ | $Fe_{1.03(1)}Te_{0.91(1)}Se_{0.09(1)}$ | $Fe_{1.04(2)}Te_{0.91(1)}Se_{0.09(1)}$ |
| Lattice parameters (pm) | a = 381.91(2)  c = 624.77(5) | a = 382.00(2)  c = 625.07(4) | a = 381.32(2)  c = 624.75(8) |
| Cell volume (nm³) | 0.09113(1) | 0.09121(1) | 0.09084(2) |
| **Atomic parameters** | | | |
| Fe1  2b (¾, ¼, 0) | | | |
| Fe2  2c (¼, ¼, z) | z = 0.711(3) | z = 0.69(1) | z = 0.710(4) |
| Te   2c (¼, ¼, z) | z = 0.281(1) | z = 0.281(1) | z = 0.281(1) |
| Se   2c (¼, ¼, z) | z = 0.239(4) | z = 0.227(8) | z = 0.233(5) |
| **Atomic distances and angles** | | | |
| Fe1-Se | 2.43(1) | 2.38(3) | 2.39(2) |
| Fe1-Te | 2.60(1) | 2.59(1) | 2.59(1) |
| Fe1-Fe2 | 2.63(1) | 2.70(1) | 2.63(2) |
| Fe1-Fe1 | 2.70(1) | 2.70(1) | 2.70(1) |
| Se-Fe-Se | 103.9(9) | 106.8(2) | 105.3(1) |
| | 112.3(5) | 110.8(9) | 111.6(6) |
| Se-Fe-Te | 99.3(5) | 100.8(9) | 100.0(6) |
| | 114.7(3) | 113.8(5) | 114.3(3) |
| Te-Fe-Te | 94.7(1) | 94.8(1) | 94.7(1) |
| | 117.3(1) | 117.3(1) | 117.3(1) |